\documentclass[conference]{IEEEtran}
\IEEEoverridecommandlockouts
\usepackage{cite}
\usepackage{amsmath,amssymb,amsfonts}
\usepackage{algorithmic}
\usepackage{graphicx}
\usepackage{textcomp}
\usepackage{xcolor}
\usepackage{algorithmic}
\usepackage{graphicx}
\usepackage{textcomp}
\usepackage{xcolor}
\usepackage{booktabs}
\usepackage{tabularx}
\usepackage{hyperref}
\usepackage{multirow}
\usepackage{amssymb}
\usepackage{multirow}
\usepackage{pifont}% http://ctan.org/pkg/pifont
\usepackage{xcolor}
\newcommand{\cmark}{\ding{51}}%
\newcommand{\xmark}{\ding{55}}%

\newtheorem{theorem}{Theorem}

\def\BibTeX{{\rm B\kern-.05em{\sc i\kern-.025em b}\kern-.08em
    T\kern-.1667em\lower.7ex\hbox{E}\kern-.125emX}}

\title{Infant Cry Detection Using Causal Temporal Representation\thanks{$^\ast$Equal contribution. \\ 
$^\dagger$This project was carried out as part of an internship at MBZUAI.}}

\author{\IEEEauthorblockN{Minghao Fu$^{1,2}$$^\ast$, Danning Li$^{3,\dagger}$$^\ast$, Aryan Gadhiya$^\dagger$, Benjamin Lambright$^\dagger$,   {Mohamed Alowais}$^\dagger$,  \\ Mohab Bahnassy$^\dagger$, {Saad El Dine Elletter}$^\dagger$, {Hawau Olamide Toyin}$^1$, {Haiyan Jiang}$^{1,2}$ \\ {Kun Zhang}$^{1,2}$, {Hanan Aldarmaki}$^1$} \\
\IEEEauthorblockN{$^1$\textit{Mohamed Bin Zayed University of Artificial Intelligence} \\
$^2$\textit{Cradle AI} \hspace{5pt} $^3$\textit{McGill University}}
}

\begin{document}

% \and
% \and
% \and
% \and
% \and

\maketitle

\begin{abstract}
This paper addresses a major challenge in acoustic event detection, in particular 
infant cry detection in the presence of other sounds and background noises: the lack of precise annotated data. We present two contributions for supervised and unsupervised infant cry detection. The first is an annotated dataset for cry segmentation, which enables supervised models to achieve state-of-the-art performance. 
Additionally, we propose a novel unsupervised method, Causal Representation Spare Transition Clustering (CRSTC), based on causal temporal representation, which helps address the issue of data scarcity more generally. 
By integrating the detected cry segments, we significantly improve the performance of downstream infant cry classification, highlighting the potential of this approach for infant care applications. 
\end{abstract}


\section{Introduction}
\label{sec:intro}
%Each year, about 130 million babies are born worldwide. 
Caring for newborns, especially for first-time parents, is a significant challenge. %While advice from other parents and books can be helpful, these sources often fall short when addressing practical issues. 
One of the main difficulties is understanding the meaning of infant cries. %, as crying is the primary way babies communicate.
In response, numerous studies have emerged to address this problem. Early research showed that trained adult listeners could differentiate between types of cries. For example, \cite{wasz1964identification} first identified four types of cries (pain, hunger, birth, and pleasure) by training nurses to recognize them. However, at best, the accuracy of trained nurses is only up to 33.09\%.
 %Methods such as \cite{mukhopadhyay2013evaluation, Sachin2017GPUBD} developed machine learning models to understand infant cries, paving the way for intelligent robotic caregivers in the future. 
Beyond recognizing infants' daily needs, disease prediction is another critical task in infant cry research. This includes identifying conditions such as hypo-acoustic, asphyxia, hypothyroidism, hyperbilirubinemia, and cleft palate. Several studies \cite{saraswathy2012automatic} have focused on classifying infant cries to aid diagnosis.

Thanks to recent advances in machine learning, automated approaches have become more feasible for understanding infant cries ~\cite{petroni1995classification}. However, recent work \cite{ji2021review} points out that while machine learning methods achieve impressive results on public datasets such as the Baby Chillanto database \cite{reyes2004classification}, their performance drops significantly on real-world data. %This is mainly due to two factors: 1) data quality and 2) the unpredictability of cry segments. 
Most of previously-explored machine learning methods for this task were modeled as binary audio classification at the level of a whole audio segment that may contain just a brief instance of a cry within a longer segment of background sounds. This is due to the nature of existing benchmarks and datasets that do not contain fine-grained labeling. This makes them insufficient for real-time baby cry detection; real-world recordings often contain not just infant cries but also background noise like parents talking, TV or phone sounds, and environmental noises, which complicates the analysis and leads to inferior results. In \cite{yao2022infantcryingdetectionrealworld}, a dataset was collected in a real-world environment and annotated for cry detection. %However, the dataset is not accessible. 
%Their model achieves an F1 score of 0.613, and 
They report that a model trained on this dataset generalizes better to real-world situations, compared to models trained using in-lab data. The dataset, however, is not publicly accessible at the time of writing.  %In this work, we attempt to address three questions: (i) how to deal with the lack of baselines and publicly available annotated datasets for baby cry detection, (ii) the lack of knowledge on how the underlying factors of audio decide type of sounds which make the algorithm on automatically detecting temporal segmentation be hard. \todo{please rephrase question ii} (iii) does accurate temporal cry detection benefit downstream tasks like baby cry classification. %Hanan: I'm commenting this out for now as it is redundant. 

In this work, We aim to develop robust models for accurate infant cry detection within audio segments of various noises, which can benefit downstream tasks and improve caregiver response times. Identifying relevant audio features in domestic environments is challenging due to diverse background sounds and the limited availability of high-quality annotated data for specific cases like baby cries. We address this issue through manual annotation and data augmentation techniques, improving baby cry analysis models by reducing noise during cry interval extraction. 
In addition, as the acquisition of annotated data is both costly and challenging, we propose a viable alternative using unsupervised methods to detect infant cry segment boundaries by approximating the underlying data-generating process. A particularly promising approach involves causal temporal representation learning \cite{scholkopf2021toward}, given that infant cries constitute a typical time-series exhibiting significant nonstationarity~\cite{ji2021review,Song2024Causal}.
Causal representation learning has been extensively utilized in the analysis of temporal dependencies within scientific discovery \cite{Battaglia2013SimulationAA, Boussard2023TowardsCR, Subbaswamy2018CounterfactualNP}, 
improving medical diagnosis~\cite{Richens2020ImprovingTA} and machine learning research~\cite{Schulam2017ReliableDS}. 
Its identifiability offers a theoretical assurance that the learned representation accurately reflects the true data-generating process, subject to minor indeterminacies.

Overall, our contributions in this paper are three-fold: (1) We present the first \textbf{annotated dataset } specifically designed for baby cry detection, and build a comprehensive supervised learning framework for data preprocessing, training and evaluation, 
(2) we propose a \textbf{novel method} named \textbf{C}ausal Representation \textbf{S}parse \textbf{T}ransition Clustering (\textbf{CRSTC}) for unsupervised acoustic event detection, and 
(3) We empirically demonstrate the positive effect of accurate infant cry temporal detection on the performance of \textbf{downstream infant cry analysis tasks}. 

The annotated dataset and code used for our experiments are available at \url{https://github.com/PeterIsDanning/Infant-Cry-Detection-by-CRSTC}.

% TODO: INTERNS CONFIRM THE DETAILS FOR MFCC, THIS DOESN'T READ WELL. 

% \noindent \textbf{MFCC} For the MFCC, the original audio was converted first, then we concatenated all of the MFCCs together, and took 300 frame segments as samples to be inputted into the model.

% \noindent \textbf{Mel-Spectrogram} 
% We padded all audio samples to 8 seconds and then extracted the mel-spectrograms using librosa. 

% Both the audio and labels were segmented into 50ms segments, yielding 160 segments to be inputted into the model per audio sample.

%\section{Our Method}
%\label{sec:typestyle}

%\subsection{Supervised Learning Method}

%\subsubsection{Method}

%\paragraph{Feature Extraction}

\section{Unsupervised Infant Cry Detection}

The absence of an annotated segmented datasets for infant cry detection considerably impedes accurate temporal identification. Moreover, acquiring annotated segmented data for medical applications involves prohibitive costs. 
To alleviate the need for manual labeling, we propose an unsupervised learning methodology for detecting acoustic events from unlabeled data using causal temporal representations that align with the underlying data generation process. We demonstrate the application of this approach for infant cry detection.  %without relying on expert knowledge or predefined characteristics. 

\subsection{Causal Temporal Representation Learning} \label{sec:ctrl}
\begin{figure}[htbp]
\centering
    \includegraphics[width=\linewidth]{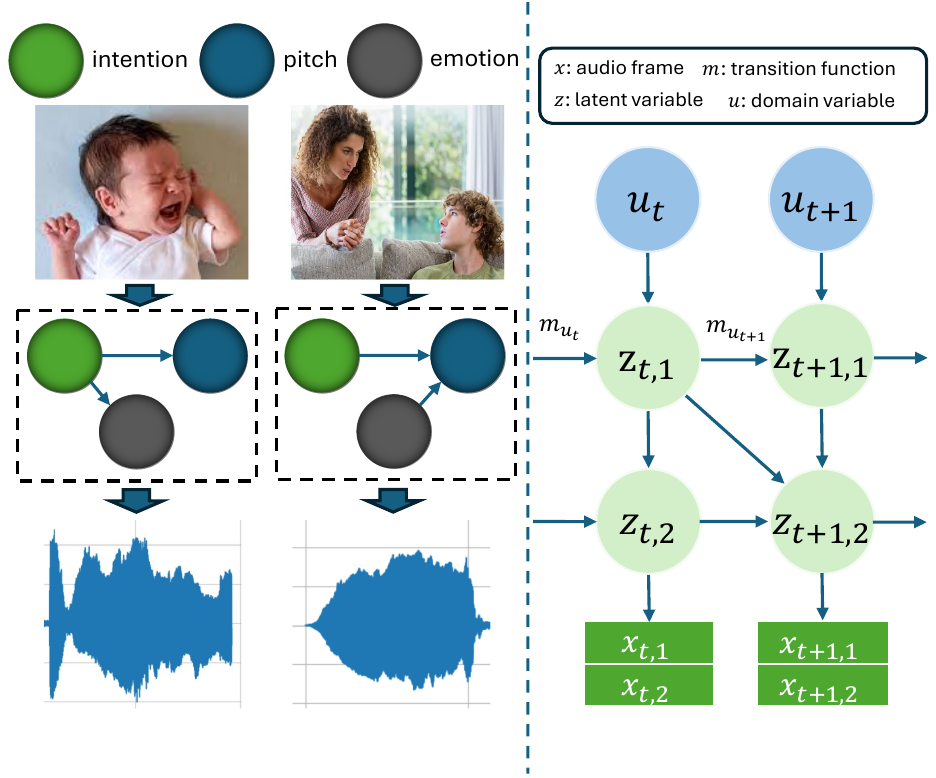}
\caption{Illustration of graphical model for nonstationary temporal data with unobserved domain variables $u_t$}
    \label{fig:DGP of baby cry}
\end{figure}

We interpret that the observed audio data are generated by causally related latent temporal processes with unknown domain variables as shown in ~\ref{fig:DGP of baby cry}. An audio segmentation after \(\mathcal{D} = \{\mathbf{x}_t\}_{t=1}^T\), where the $t$-th frame \(\mathbf{x}_t \in \mathcal{X} \) is produced from causally related, time-delayed \textit{latent representation} \(\mathbf{z}_t \in \mathbb{R}^n\) through an invertible mixing function \(g\):
\begin{equation}
    \mathbf{x}_t = g(\mathbf{z}_t).
\end{equation}
$\mathbf{z}_t$ of baby cry audio signals encapsulates factors crucial for generating these signals, capturing key characteristics like pitch, tone, and intensity variations for accurate understanding. The \(i\)-th component of \(\mathbf{z}_t\) is generated by the \(i\)-th component of \(m_{u_t}\) via a Markov Process:
\begin{equation}
    z_{t,i} = m_{i, u_t} \left(\{z_{t-1, j} \mid z_{t-1, j} \in \mathbf{Pa}(z_{t,i})\}, \epsilon_{t,i} \right),
\end{equation}
Where \(\mathbf{Pa}(z_{t,i})\) represents latent factors at step \(t-1\) influencing \(z_{t,i}\), the domain variable \(u_t \in \mathcal{U}\) becomes relevant. This aligns with humans' ability to recognize domain shifts with enough transition variation, as shown in tasks like video action segmentation and recognition \cite{xu2024efficient}. In baby cry audio, domain variables are different environments. In nonstationary settings, causal dependency graphs change across domains. We denote this transition function as \(m_{u_t} \in \mathcal{M}\), assuming each \(i\)-th component of \(z_t\) is mutually independent, conditioned on \(z_{t-1}\) and \(u_t\). Noise terms \(\epsilon_{t,i}\) are also assumed to be mutually independent across dimensions and over time.
% Figure~\ref{fig:DGP of baby cry} illustrates the graphical model for this setting.
% ridley2022transformers, 

\begin{figure*}
\centering
    \includegraphics[width=1.0\textwidth]{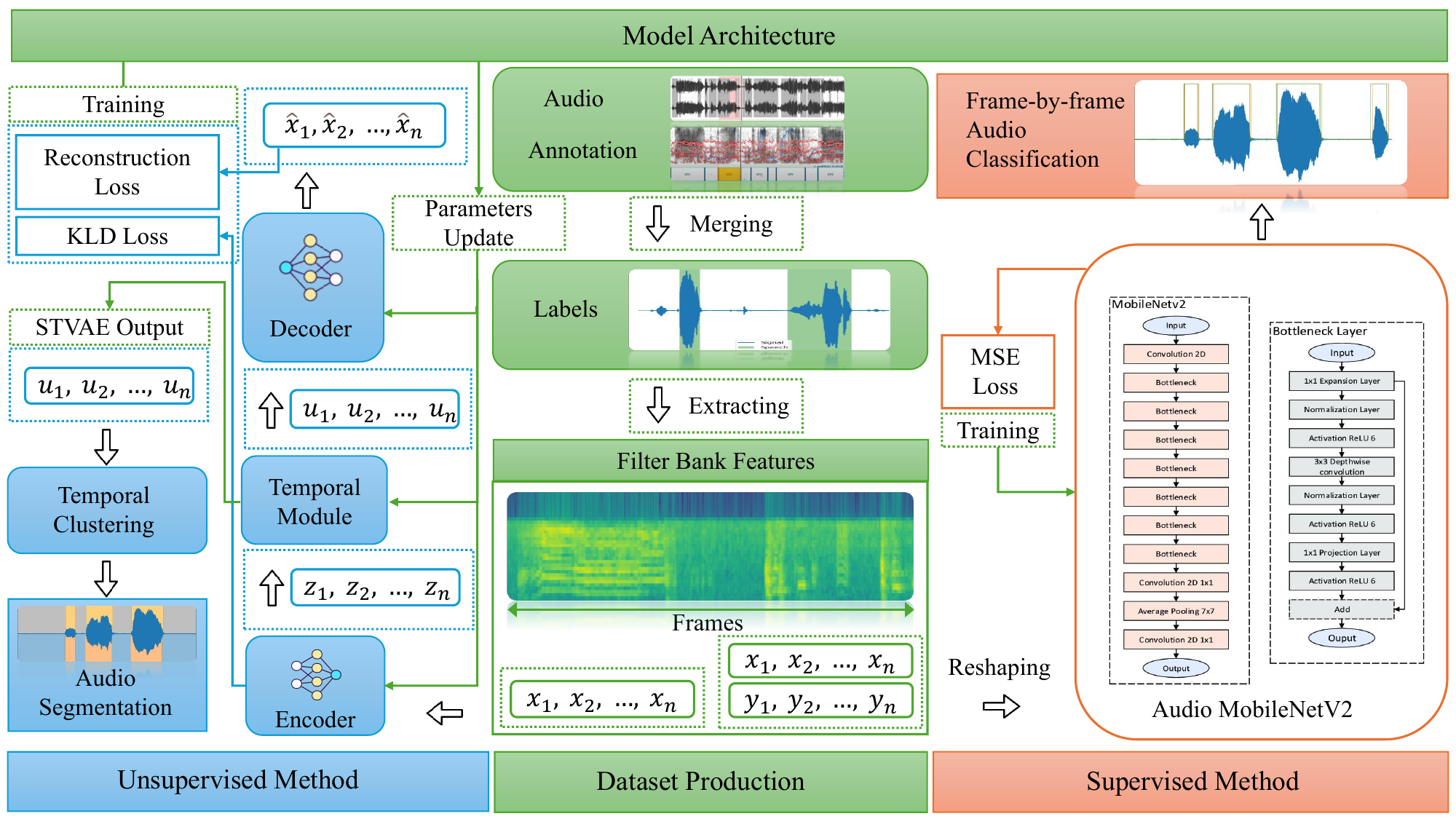}
\caption{Sparse transition variational autoencoder with temporal clustering}
    \label{fig:Model structure}
\end{figure*}

\subsection{Sparse Transition Clustering}
We hypothesize that the observed audio data arise from causally related latent temporal processes governed by unknown domain-specific variables. We employ a modified Variational Autoencoder (VAE), the \textbf{Sparse Transition Variational Autoencoder} (ST-VAE), which we design to effectively encode these unknown variables. Figure~\ref{fig:Model structure} illustrates the comprehensive architecture of this model. 

The ST-VAE architecture enhances the traditional VAE with a temporal module. The encoder, with multiple perception layers, encodes audio features into a causal temporal latent distribution. The decoder, also with multiple perception layers, reconstructs audio from latent variables and computes reconstruction loss to update encoder parameters for precise latent distribution. The temporal module, using LSTM layers, models transitions between latent variables over time. After optimization, domain-specific variables are identified from the sparse transition module.

We employ clustering techniques for temporal segmentation of the transition variables, which assigns each frame to a cluster based on its similarity to other frames, resulting in classifying different latent temporal patterns in the audio. This process has the potential to reveal subtle distinctions between different audio classes, including baby cries. The automatic or data-driven determination of cluster number ensures flexibility and adaptability to the specific characteristics of the audio data, enhancing the robustness of our approach. 

\subsection{Theoretical Guarantee}
Note that domain variables have a unique solution up to label swapping if observational equivalence (Section~\ref{sec:ctrl}), which are shown as follows.

\begin{theorem}
(\textbf{Identifiability of Domain Variables} (directly from Thm. 1 in ~\cite{Song2024Causal}))
\label{thm: identifiability of C}
Suppose that the dataset \(\mathcal{D}\) is generated from a nonstationary data generation process. Suppose the following assumptions hold:
\begin{itemize}
    \item \label{as: Mechanism Separability}\underline{(Mechanism Separability)} There exists a ground truth mapping \(\mathcal{C}: \mathcal{X} \times \mathcal{X}  \to \mathcal{U}\) that determines the real domain indices, i.e., \(u_t = \mathcal{C}(\mathbf{x}_{t-1}, \mathbf{x}_{t})\).
    \item \label{as: Mechanism Sparsity} \underline{(Mechanism Sparsity)} The estimated transition complexity on data \(\mathcal{D}\) is less than or equal to ground truth transition complexity, i.e., \(\mathbb{E}_{\mathcal{D}} | \widehat{\mathcal{M}}_{\hat{u}} | \leq \mathbb{E}_{\mathcal{D}}  | \mathcal{M}_{u} |\).\label{eq: Mechanism Sparsity}
    
    \item \label{as: Mechanism Variability}\underline{(Mechanism Variability)} Mechanisms are sufficiently different. For all $u\neq u'$, $\mathcal{M}_{u} \neq \mathcal{M}_{u'}$ i.e. there exists index $(i,j)$ such that \(\left[\mathcal{M}_{u}\right]_{i,j} \neq \left[\mathcal{M}_{u'}\right]_{i,j}\).
\end{itemize}
Then the domain variables $u_t$ are identifiable up to label swapping.
\end{theorem}
\paragraph{Explanation} According to the theoretical analysis, if the assumptions are satisfied—loosely speaking, the mechanisms across different domains (e.g., baby cries and other sounds) being sufficiently distinct—it follows that domain labels can be uniquely determined from the audio data. Further discussion on this assumption can be found in ~\cite{Song2024Causal}.
% \begin{equation}
%     \begin{aligned}
%             p_{\hat{m}, \hat{u}, \hat{p}, \hat{e}, \hat{g}}\left( \left\{ \mathbf{x}_t \right\}_{t=1}^{T} \right) = p_{m, u, p, e, g}\left( \left\{ \mathbf{x}_t \right\}_{t=1}^{T} \right) \\ \implies \hat{u}_t = \sigma(u_t), \forall t \in \{1, 2, \dots, T\}  
%     \end{aligned}
% \end{equation}

\section{Data Annotation}
\label{sec:dataset}

We select the Donate A Cry~\cite{Gveres_2015} dataset as our main training dataset for baby cries, which including 457 samples averaging 7.72 seconds in length from the range (6.95, 8.03) seconds. The cry samples consist of 276 male and 181 female infants, and a large majority were less than 24 months old. The samples include baby cries alongside other background noises, and typically used without finer segmentation. %, which ensure its domain distribution focus on baby with little arg.

We manually annotated this data using \href{https://www.fon.hum.uva.nl/praat/}{Praat} for accurate cry detection, where annotators identify precise start and end times of infant cries within the input audio. Our data was annotated independently by three different annotators per sample, and six annotators in total. We experimented with different aggregation methods. The simplest approach was to take a `majority vote' for each point in the .wav file as each point is annotated in one of two classes: cry, no cry. The aforementioned approach can be found in our repository.
After annotation and aggregation, each input audio was padded to 8 seconds, then divided into 160 intervals of 0.05 seconds used in all models for training and testing. We randomly split the data into 80\%/20\% for train/test sets. \\

\section{Experiments}
%Discuss supervised models and their exprimental settings here

\subsection{Metrics}
We use frame-based and event-based metrics for comprehensive evaluation:

\noindent\textbf{Frame-Based Accuracy:} this metric is the simplest and evaluates the predictions on each frame independently. It quantifies the proportion of frames for which the model generates correct predictions.

\noindent\textbf{Frame-Based F1:}
In instances where audio recordings exhibit a significant disparity in the prevalence of positive and negative classes (such as baby cry versus non-baby cry segments), a conventional accuracy metric may not sufficiently reflect the model's true performance. We utilize the frame-based F1 score, which offers a balanced evaluation of both positive and negative predictions at the level of frames. 

\noindent\textbf{Event-Based F1:} 
%Within the domain of event detection, which centers on the precise identification of temporal segments within a sequence, the conventional F-score employed in classification tasks falls short in addressing the temporal intricacies inherent in audio event detection, necessitating a tailored approach. 
Frame-level evaluation falls short in addressing the temporal dependencies inherent in audio event detection. 
Drawing inspiration from \cite{Mesaros2016_MDPI}, we use the event-based F1-score, a metric that %explicitly accounts for the unique challenges posed by event detection, particularly the imperative of
incorporates the temporal alignment between predicted and ground truth events. %This metric operates on an event-by-event basis, acknowledging that, unlike traditional classification where each instance receives a single label, event detection entails discerning temporal segments within a sequence. Consequently,
A predicted event is classified as a true positive when it demonstrates sufficient overlap with a corresponding ground truth event.

\noindent\textbf{Event-Based IOU:} Intersection Over Union (IOU) is traditionally used for evaluating object detection to gauge the spatial precision of bounding box predictions. We apply this metric to audio segmentation. This metric quantifies the degree of alignment between predicted event segments and their corresponding ground truth segments.

% \begin{table*}[htb]
% \centering
% \caption{Model Evaluation of Supervised and Unsupervised Methods}\label{tab:evaluation}
% \small % Reduce font size
% \scalebox{0.9}{
% \begin{tabular}{@{} lccccc @{}}
% \toprule
% \multirow{2}{*}{\centering Models} & \multirow{2}{*}{\centering Supervised} & \multicolumn{2}{c}{Frame-Based} & \multicolumn{2}{c@{}}{Event-Based} \\ 
% \cmidrule(lr){3-4} \cmidrule(lr@{}){5-6}
%  & & F1 & Accuracy & F1 & IOU \\ 
% \midrule
% \textit{Ours} (BiLSTM \cite{Huang2015BidirectionalLM})   & \cmark    & 0.93      & 0.92      & 0.85      & \textbf{0.77} \\
% \textit{Ours} (Transformer \cite{Vaswani2017AttentionIA})   & \cmark  & 0.84 & 0.85 & 0.92 & 0.23 \\
% \textit{Ours} (AudioMobileNetV2 \cite{Sandler2018MobileNetV2IR}) & \cmark  & \textbf{0.94} & \textbf{0.94} & \textbf{0.94} & 0.60 \\
% FDY-CRNN \cite{nam2022frequency} & \cmark & 0.83 & 0.87 & 0.56 & 0.52 \\ 
% MDFD-SED \cite{nam2024pushing} & \cmark & 0.76 & 0.80 & 0.61 & 0.73 \\ 
% \textit{Ours} (CRSTC, cluster type=KMeans,k=2) & \xmark & 0.88 & 0.90 & 0.87 & 0.52 \\
% \textit{Ours} (CRSTC, cluster type=KMeans,k=4) & \xmark & 0.78 & 0.70 & 0.69 & 0.58 \\
% \textit{Ours} (CRSTC, cluster type=KMeans,k=6) & \xmark & 0.72 & 0.67 & 0.55 & 0.55 \\
% \textit{Ours} (CRSTC, cluster type=Mean-Shift) & \xmark & 0.75 & 0.74 & 0.64 & 0.41 \\
% \textit{Ours} (CRSTC, cluster type=Bisecting,k=2) & \xmark & 0.87 & 0.88 & 0.62 & 0.30 \\
% \bottomrule
% \end{tabular}
% }
% \normalsize % Restore normal font size
% \end{table*}

\subsection{Results}
\textbf{Supervised Learning Baselines}
We experimented with the BiLSTM, %~\cite{Huang2015BidirectionalLM}, 
MobileNetV2~\cite{Sandler2018MobileNetV2IR}, and Transformer %~\cite{Vaswani2017AttentionIA} 
architectures for supervised training. We also compare with several audio event detection methodologies, including FDY-CRNN \cite{nam2022frequency} and MDFD-SED \cite{nam2024pushing}, employing their model structures and backpropagation strategies. The choice of feature extraction techniques was guided by their demonstrated superiority in frame-based accuracy. All models were trained using our manually labeled cry detection dataset described in section \ref{sec:dataset}. The results are shown in Table~\ref{tab:evaluation}. 
These results illustrate the effectiveness of both BiLSTM and AudioMobileNetV2 compared to baselines. %, which is evidenced by its strong performance at both the frame and event levels. 
%The enhancements in our methods compared to conventional audio event detection literature indicate the suitability of our framework for supervised temporal detection tasks. %\todo{which features were used here? filterbanks similar to the unsupervised model?}

\begin{table}
\centering
\setlength{\tabcolsep}{3pt}
\caption{Evaluation of Supervised and Unsupervised Models}\label{tab:evaluation}
\small % Reduce font size
\resizebox{\linewidth}{!}{
\begin{tabular}{lccccc}
\toprule
\multirow{2}{*}{Models} & \multirow{2}{*}{Setting} & \multicolumn{2}{c}{Frame-Based} & \multicolumn{2}{c@{}}{Event-Based} \\ 
\cmidrule(lr){3-4} \cmidrule(lr){5-6}
 & & F1 & Accuracy & F1 & IOU \\ 
\midrule
Baseline: FDY-CRNN \cite{nam2022frequency} & supervised & 0.83 & 0.87 & 0.56 & 0.52 \\ 
Baseline: MDFD-SED \cite{nam2024pushing} & supervised & 0.76 & 0.80 & 0.61 & 0.73 \\ 
\textit{Ours} (BiLSTM \cite{Huang2015BidirectionalLM})   & supervised    & 0.93      & 0.92      & 0.85      & \textcolor{red}{0.77} \\
\textit{Ours} (Transformer \cite{Vaswani2017AttentionIA})   & supervised  & 0.84 & 0.85 & 0.92 & 0.23 \\
\textit{Ours} (AudioMobileNetV2 \cite{Sandler2018MobileNetV2IR}) & supervised  & \textcolor{red}{0.94} & \textcolor{red}{0.94} & \textcolor{red}{0.94} & 0.60 \\
\textit{Ours} (CRSTC, KMeans, $k=2$) & unsupervised & \textcolor{blue}{0.88} & \textcolor{blue}{0.90} & \textcolor{blue}{0.87} & 0.52 \\
\textit{Ours} (CRSTC, KMeans, $k=4$) & unsupervised & 0.78 & 0.70 & 0.69 & \textcolor{blue}{0.58} \\
\textit{Ours} (CRSTC, KMeans, $k=6$) & unsupervised & 0.72 & 0.67 & 0.55 & 0.55 \\
\textit{Ours} (CRSTC, Mean-Shift) & unsupervised & 0.75 & 0.74 & 0.64 & 0.41 \\
\textit{Ours} (CRSTC, Bisecting, $k=2$) & unsupervised & 0.87 & 0.88 & 0.62 & 0.30 \\
\bottomrule
\end{tabular}
}
\normalsize % Restore normal font size
\end{table}

% \begin{table*}[htb]
% \centering
% \caption{Model Evaluation}\label{tab:evaluation}
% \small % Reduce font size
% \begin{tabularx}{\linewidth}{lCCCCCC}
% \toprule
% Models & Using Labeled Data & Frame-Based F1 & Event-Based F1 & Frame-Based Accuracy & Event-Based IOU\\ 
% \midrule
% \textit{Ours} (BiLSTM \cite{Huang2015BidirectionalLM})   & \cmark    & 0.76                 & 0.64                 & 0.83     \\
% \textit{Ours} (MobileNetV2 \cite{Sandler2018MobileNetV2IR}) & \cmark  & \textbf{0.87}        & \textbf{0.87}        & \textbf{0.92} \\
% \textit{Ours} (Transformer \cite{Vaswani2017AttentionIA})   & \cmark  & 0.82                 & 0.82                 & \textbf{0.92} \\
% FDY-CRNN \cite{nam2022frequency} & \cmark & 0.83 & 0.56 & 0.87 & 0.52 \\ 
% MDFD-SED \cite{nam2024pushing} & \cmark & 0.76 & 0.61& 0.80 & 0.73 \\ 

% \textit{Ours} (CRSTC, cluster type=KMeans,k=2) & \xmark & 0.92 & 0.79 & 0.95 & 0.77 \\
% \textit{Ours} (CRSTC, cluster type=KMeans,k=4) & \xmark & 0.90 & 0.74 & 0.91 & 0.80 \\
% \textit{Ours} (CRSTC, cluster type=KMeans,k=6) & \xmark & 0.67 & 0.58 & 0.91 & 0.68 \\
% \textit{Ours} (CRSTC, cluster type=Mean-Shift) & \xmark & 0.72 & 0.63 & 0.85 & 0.67\\
% \textit{Ours} (CRSTC, cluster type=Bisecting,k=2) & \xmark & 0.93 & 0.77 & 0.94 & 0.78\\
% \bottomrule
% \end{tabularx}
% \normalsize % Restore normal font size
% \end{table*}

\begin{figure}[t]
    \centering
    \includegraphics[width=85mm]{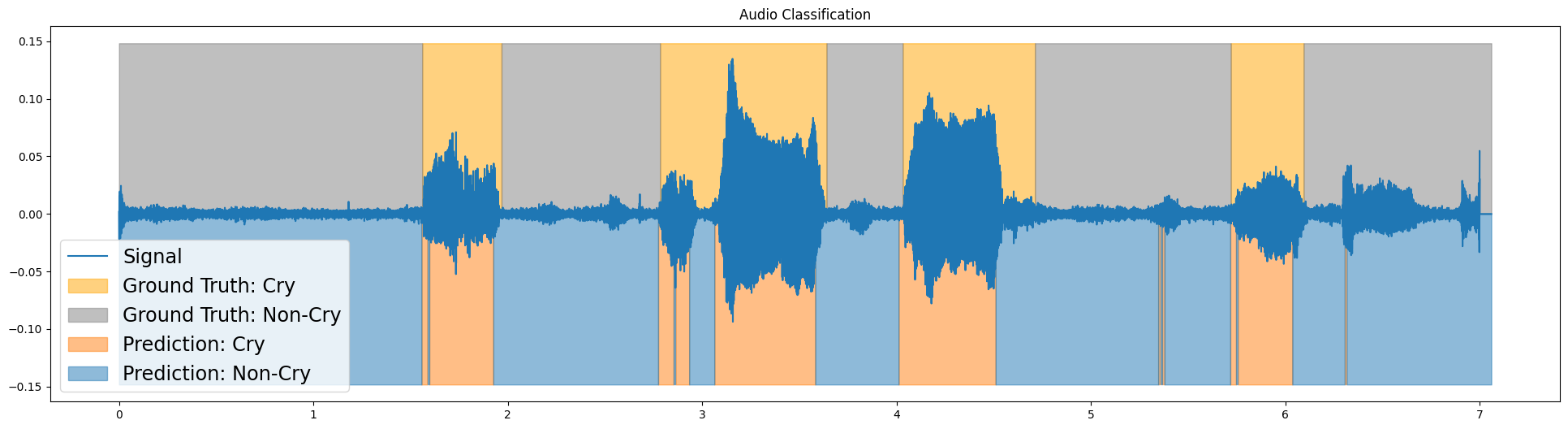}
    \includegraphics[width=85mm]{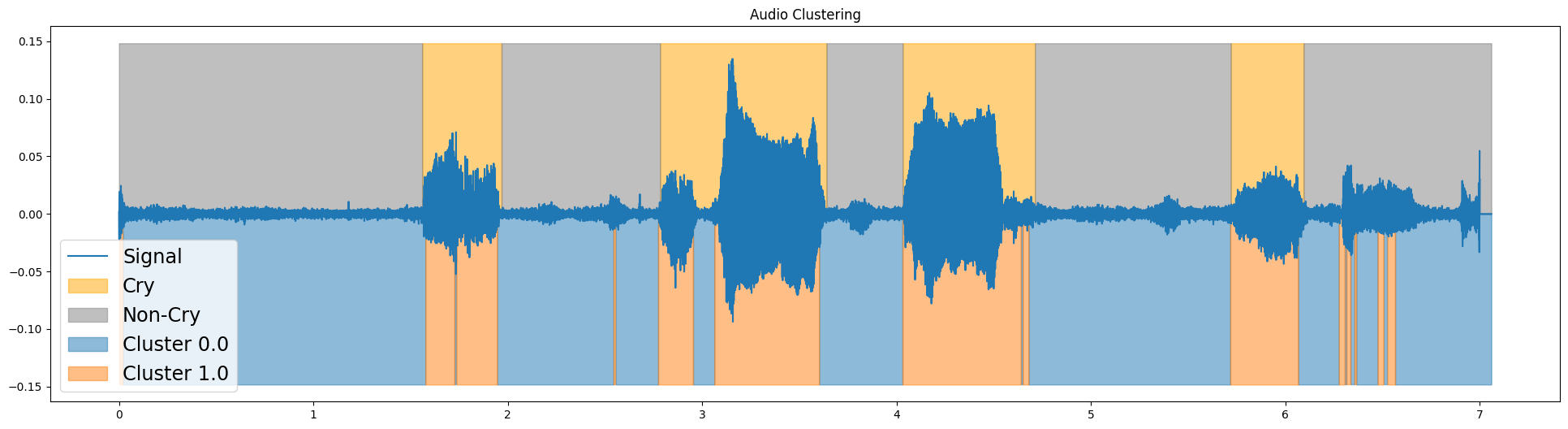}
    \caption{Comparison of Supervised (top) and Unsupervised (bottom) infant cry detection model predictions.}
    \label{fig:combined}
\end{figure}

\textbf{Unsupervised Learning}
The performance of our proposed unsupervised learning algorithm is also shown in Table~\ref{tab:evaluation} with difference clustering settings. These results demonstrate competitive performance of the proposed method, even exceeding some of the supervised baselines. %and even outperforms the supervised method. 
%This highlights a key limitation of supervised learning—when the data scale is limited, the model's ability to generalize is restricted by its dependence on labeled data. In contrast, unsupervised learning can leverage the causal structure and patterns within the data itself, leading to better results in scenarios with insufficient labeled examples.
A visual illustration of the boundary predictions of supervised and unsupervised methods is illustrated in Figure ~\ref{fig:combined}. %~\ref{fig:unsupervised}.

% \begin{figure}[ht!]
%     \centering
%     \includegraphics[width=80mm]{figures/supervised_output.png}
%     \caption{Supervised Method Prediction \label{fig:supervised}}
%     \end{figure}

%     \begin{figure}[ht!]
%     \centering
%     \includegraphics[width=80mm]{figures/unsupervised_output.png}
%     \caption{Unsupervised Method Prediction \label{fig:unsupervised}}
%     \end{figure}
    
\subsection{Improvement on Downstream Tasks}
To test the effectiveness of accurate detection on downstream applications, we perform a comparative evaluation of various baby cry classification datasets and models with and without the detection of cry segments. We utilize publicly available cry classification datasets ~\cite{reyes2004classification,Gveres_2015} as well as a proprietary dataset ~\cite{CradleAI}. We select the AudioMobileNetV2 as the cry detection model to yield the best results. To assess detection efficiency, we employ widely recognized classification methods ~\cite{Maghfira2020InfantCC,Huckvale2018NeuralNA,turan2018monitoring} alongside our private state-of-the-art approach ~\cite{CradleAI}. 

As evidenced in Table~\ref{tab:downstream tasks}, accurate baby cry detection significantly enhances downstream classification performance across all evaluated models, indicating its effectiveness and broad applicability of accurate acoustic event detection in downstream applications.

\begin{table}[htbp] 
\label{tab:downstream tasks}
\centering
\caption{Improvements on Baby Cry Classification} 
\small % Reduce font size
\renewcommand{\arraystretch}{0.35} % Adjust row spacing
\scalebox{0.9}{ % Adjust table scale
\begin{tabular}{@{}l| lccc @{}}
\toprule
Dataset & Models & Detection & F1 & Acc (\%) \\ 
\midrule
\multirow{8}{*}{BabyChillanto~\cite{reyes2004classification}} & \multirow{2}{*}{CNN-RNN~\cite{Maghfira2020InfantCC}} & \xmark & 0.76 & 85.76 \\ 
& & \cmark & 0.81 & 90.30 \\
\cmidrule{2-5}
& \multirow{2}{*}{LSTM~\cite{Huckvale2018NeuralNA}} & \xmark & 0.73 & 82.44 \\
&  & \cmark & 0.77 & 83.28 \\
\cmidrule{2-5}
& \multirow{2}{*}{Capsule~\cite{turan2018monitoring}} & \xmark & 0.80 &  89.72 \\
& & \cmark & 0.82 & 91.70 \\
\cmidrule{2-5}
& \multirow{2}{*}{BabbleNet~\cite{CradleAI}} & \xmark & 0.89 &  95.72 \\
& & \cmark & 0.92 & 96.06 \\
\midrule
\multirow{8}{*}{Donate A Cry~\cite{Gveres_2015}} & \multirow{2}{*}{CNN-RNN~\cite{Maghfira2020InfantCC}} & \xmark & 0.71 & 80.31 \\ 
& & \cmark & 0.80 & 87.30 \\
\cmidrule{2-5}
& \multirow{2}{*}{LSTM~\cite{Huckvale2018NeuralNA}} & \xmark & 0.73 & 77.53 \\
&  & \cmark & 0.79 & 81.62 \\
\cmidrule{2-5}
& \multirow{2}{*}{Capsule~\cite{turan2018monitoring}} & \xmark & 0.75 &  85.64 \\
& & \cmark & 0.85 & 89.98 \\
\cmidrule{2-5}
& \multirow{2}{*}{BabbleNet~\cite{CradleAI}} & \xmark & 0.87 &  94.72 \\
& & \cmark & 0.91 & 95.85 \\
\midrule
\multirow{8}{*}{BabbleNet~\cite{CradleAI}} & \multirow{2}{*}{CNN-RNN~\cite{Maghfira2020InfantCC}} & \xmark & 0.67 & 78.23 \\ 
& & \cmark & 0.75 & 84.12 \\
\cmidrule{2-5}
& \multirow{2}{*}{LSTM~\cite{Huckvale2018NeuralNA}} & \xmark & 0.70 & 75.48 \\
&  & \cmark & 0.73 & 80.79 \\
\cmidrule{2-5}
& \multirow{2}{*}{Capsule~\cite{turan2018monitoring}} & \xmark & 0.72 & 82.47 \\
& & \cmark & 0.80 & 86.62 \\
\cmidrule{2-5}
& \multirow{2}{*}{BabbleNet~\cite{CradleAI}} & \xmark & 0.83 & 90.12 \\
& & \cmark & 0.88 & 93.06 \\
\bottomrule
\end{tabular}
}
\normalsize % Restore normal font size
\end{table}
\section{Conclusion}
\label{sec:print}
In this work, we contributed two methods to improve the performance of infant cry detection: an annotated dataset for precise boundary detection, and an unsupervised learning method that can be used in the absence of supervised data. Both approaches showed promising results, pushing the state-of-the-art in the supervised setting, and providing competitive results in the unsupervised case. We also showed how accurate detection improves downstream infant cry classification consistently across different models and datasets. It can serve as a useful tool for research applications in baby care, contributing effectively to societal benefits. 
\newpage
% \bibliographystyle{IEEEbib}
% \bibliography{refs}

\end{document}